\documentclass[12pt]{article}
\usepackage{graphicx,amsmath,float}
\usepackage{units}

\newlength{\dinwidth}
\newlength{\dinmargin}
\def\lapproxeq{\lower .7ex\hbox{$\;\stackrel{\textstyle                                                    
<}{\sim}\;$}}                                                    
\def\gapproxeq{\lower .7ex\hbox{$\;\stackrel{\textstyle                                                    
>}{\sim}\;$}}                                                    
\def\be{\begin{equation}}                                                    
\def\ee{\end{equation}}                                                    
\def\bea{\begin{eqnarray}}                                                    
\def\eea{\end{eqnarray}}

\def\sh{\hat s}
\def\sh2{{\hat s}^2}

\begin{document}
                                                    
\titlepage                                                    
\begin{flushright}                              CERN-TH-2019-137 \\                       
IPPP/19/65  \\  
LTH 1213 \\                             \vspace{0.3cm}                 
\today \\                                                    
\end{flushright} 
\vspace*{0.5cm}
\begin{center}                                                    
{\Large \bf How to include exclusive $J/\psi$ production data}\\
\vspace{0.5cm}
{\Large \bf   in global PDF analyses}\\
\vspace*{1cm}
                                                   
C.A. Flett$^a$, S.P. Jones$^b$, A.D. Martin$^c$, M.G. Ryskin$^{c,d}$ and T. Teubner$^a$\\                                                    
                                                   
\vspace*{0.5cm}                                                    
$^a$  Department of Mathematical Sciences, University of Liverpool, Liverpool, L69 3BX, U.K.\\
$^b$  Theoretical Physics Department, CERN, Geneva, Switzerland \\
$^c$ Institute for Particle Physics Phenomenology, Durham University, Durham, DH1 3LE, U.K. \\                                                   
$^d$ Petersburg Nuclear Physics Institute, NRC Kurchatov Institute, Gatchina, St.~Petersburg, 188300, Russia

\vspace*{1cm}                                                    
                                                    
\begin{abstract} 
We compare the cross section for exclusive $J/\psi$ photoproduction calculated at NLO in the collinear factorization approach with HERA and LHCb data. Using the optimum scale formalism together with the subtraction of the low $k_t$ contribution (below the input scale $Q_0$) from the NLO coefficient function to avoid double counting, we show that the existing global parton distribution functions (PDFs) are consistent with the data within their uncertainties.  This is the first time that $J/\psi$ production data at HERA were successfully described within the NLO collinear factorization framework using the PDFs of the global parton analyses.  However, at lower $x$ the uncertainties of the present global PDFs are large.  On the other hand, the accuracy of the LHCb data are rather good.  Therefore, these data provide the possibility to directly measure the gluon PDF over the very large interval of $x$, ~ $10^{-6}<x<10^{-2}$,~ at a fixed low scale.

\end{abstract}

\vspace*{0.5cm}                                                    
                                                    
\end{center}

 \section{Introduction}
 \label{sec:intro}

The parton distributions of the proton are known with good precision from global parton analyses  as long as the $x$ values are not too low.  Indeed for $x\gapproxeq 10^{-3}$ the results of the different groups \cite{NNPDF,MMHT,CT14} agree with each other quite well. However, the uncertainty in the parton distributions strongly increases as we go to lower values of $x$, especially at low scales.  This simply reflects the fact that no experimental data are used to directly probe this region.\footnote{Besides its intrinsic value, there are at least two further reasons to be interested in the behaviour of the gluon PDF at very small $x$ and low scales $\mu \sim 1.5\, \text{GeV}$.  First, recall that the distribution of gluons as $x\to 0$  governs the high-energy asymptotics of the scattering amplitude. In particular, the gluon distribution at some relatively low scale can be used as the boundary condition for the BFKL equation. This boundary condition for BFKL is needed to account for the effects of confinement.  As was shown in \cite{Kol,KLRS}, such a boundary condition replaces the BFKL cut (in the complex momentum $j$-plane) by a series of Regge poles.
At very low $x$ the boundary condition should indicate the presence of saturation effects that are needed to stop the power growth of the original BFKL amplitude.
Another motivation for obtaining a reliable gluon PDF at small $x$ is that it may be used to evaluate the production cross section of a possible new light particle at the LHC (if such a new particle exists) or to put a limit on the corresponding coupling.}

On the other hand the LHCb detector has the possibility of particle detection in the rapidity range $2<Y<4.5$. In particular the collaboration have measured the differential cross sections for open charm \cite{cc} (and bottom \cite{bb}) quark pairs, and also for exclusive $J/\psi$ (and $\Upsilon$) vector mesons \cite{LHCb}, which allow the determination of the low $x$ gluon PDF for $x \sim 10^{-5}$ or less at factorization scales $\mu_F=\sqrt{m^2_q +p^2_{T,q}}$ and $\mu_F=m_q$, respectively, where $q=c,b$ and $p_T$ is the transverse momentum of the quark. 

The differential cross sections for open $c\bar{c},~b\bar{b}$ production are determined by LHCb \cite{cc,bb} by observing $D$ and $B$ meson decays. These data are then studied to extract information about the gluon PDF at low $x$ \cite{r4,r5,r6,r7,Gauld,OMR}. Here, we may say the experimental measurement is not simple while the theory is more straightforward. In fact careful analyses, for example,~\cite{Gauld,OMR} indicate that there are serious tensions and inconsistencies in the $D$ and $B$ data, and that no conclusion about the very low $x$ behaviour of the gluon PDF is possible.  In a sense, for exclusive $J/\psi$ the opposite is true. The LHCb data are more straightforward to collect and the accuracy of the exclusive $J/\psi$ differential cross sections is much better \cite{LHCb}. However here the theory is more involved. In short there are two theoretical problems to address.
First, the corresponding cross section is not described by the usual PDFs but by the more complicated generalised parton distributions (GPDs), see \cite{GPD} for a review. Next, the NLO corrections are large and the results strongly depend on the choice of scale.

In the present paper we recall how these two problems can be solved within the conventional collinear approach  by using the Shuvaev transform \cite{Shuv}, which at small $x$ allows the calculation of the GPDs from the conventional integrated PDFs. Secondly, the strong scale dependence can be reduced by choosing a factorization scale which effectively resums the double logarithmic $\alpha_s \ln (\mu^2) \ln(1/x)$ terms (which are enhanced by the large values of ln$(1/x)$ at small $x$) and transfers them into the incoming PDFs. Finally, and most importantly, to avoid double counting, we have to subtract the low transverse momentum, $k_t$, contributions below the input scale $Q_0$ from the NLO coefficient functions, as these contributions are already included in the input PDFs.  The subtraction is of the form of a power correction which, as expected, is large.

Previously, the LHCb data for forward ultraperipheral $J/\psi$ production were successfully described in \cite{Jones} using the $k_t$ factorization framework. However, the $k_t$ factorization approach does not include the complete set of NLO corrections. Thus this approach does not allow these $J/\psi$ data to be included in the NLO global analyses based on the collinear factorization theorems.  Our formalism is based on the conventional collinear framework and includes all NLO corrections. In Section~\ref{sec:data} we show that three existing sets of PDFs (NNPDF3.0 \cite{NNPDF}, MMHT2014 \cite{MMHT}, CT14 {\cite{CT14}) taken at the optimal scale mentioned above, and convoluted with the NLO coefficient functions from which the low $k_t<Q_0$ contribution has been subtracted, give a satisfactory description of the diffractive $J/\psi$ HERA data, but vastly different predictions in the region of the LHCb $J/\psi$ data. Here, $Q_0$ is the PDF input scale.

The plan of the paper is as follows. In Section~\ref{sec:notation} we give our notation. In Section~\ref{sec:pdfs} we explain how our approach can be used to probe the PDFs. In Section~\ref{sec:scale} we demonstrate the stability of the analysis with respect to variations of the remaining scale dependence. In Section~\ref{sec:data} we show that the PDFs given by the existing global analyses agree with the $J/\psi$ exclusive photoproduction data measured at HERA \cite{HERA} and that they can be constrained at even smaller $x \sim 10^{-6}$ using LHCb ultraperipheral $J/\psi$ data. We discuss our results in Section~\ref{sec:results}} and present our conclusions in Section~\ref{sec:conclusions}.



\section{Notation \& collinear factorization}
\label{sec:notation}

The exclusive $J/\psi$ photoproduction amplitude may be written, using collinear factorization, in the form~\cite{Ivan}
\be
A = \frac{4 \pi \sqrt{4 \pi \alpha} e_q ( \epsilon_V^* \cdot \epsilon_\gamma)}{N_c} \left(\frac{ \langle O_1 \rangle_V}{m_c^3} \right)^{1/2} \int_{-1}^1 \frac{\mathrm{d}X}{X}\, \left[ C_g\left( X,\xi \right) F_g(X,\xi) + C_q(X,\xi) F_q(X,\xi) \right],
\ee
where we have suppressed the dependence on the renormalization and factorization scales, $\mu_R^2, \mu_F^2$, and on the invariant transferred momentum $t$. Here, the non-relativistic QCD (NRQCD) matrix element $\langle O_1 \rangle_V$ describes the formation of the $J/\psi$ meson with $m_c$ the charm quark mass. The quark singlet and gluon GPDs are denoted $F_q$ and $F_g$, respectively. The quark and gluon coefficient functions $C_q$ and $C_g$ are known at NLO~\cite{Ivan} and are given at tree level by
\bea
C_g^{(0)}(X,\xi) &=&  \alpha_s \frac{X}{(X-\xi + i \varepsilon)(X+\xi - i \varepsilon)} \left( \frac{2}{d-2} \right), \nonumber \\
C_q^{(0)}(X,\xi) &=& 0, \nonumber
\eea
where $d=4-2\epsilon$ is the number of space-time dimensions. 

The kinematics of the process are displayed in Fig.~\ref{fig:f2}. The partons carry momentum fractions $(X+\xi)$ and $(X-\xi)$ of the plus-component of the mean incoming/outgoing proton momenta $P=(p+p^\prime)/2$. The photon-proton centre of mass energy squared is given by $W^2=(q+p)^2$, where $q$ is the photon momentum. The asymmetry between the momentum fractions carried by the partons is parametrised by the skewness parameter,
\be
\xi = \frac{p^{+} - p^{\prime +}}{p^{+} + p^{\prime +}} = \frac{M_\psi^2}{2 W^2 - M_\psi^2}.
\ee

Due to the vanishing of the quark coefficient function at LO the process is predominantly sensitive to the gluon GPD. At LO, the gluon coefficient function is strongly peaked for $|X| \sim \xi$ and so the gluon GPD is probed close to $F_g(\xi,\xi)$. In fact, for the imaginary part of the amplitude, the LO gluon coefficient function acts as a Dirac delta function and the GPD is probed at exactly $|X|=\xi$.

\section{Connecting exclusive production to the PDFs}
\label{sec:pdfs}

Firstly, let us recall the advantage of using the exclusive $J/\psi$ LHCb data in global parton analyses in the collinear factorization scheme. It offers the possibility to probe PDFs (mainly the gluon PDF) at extremely low $x$ in a so far unexplored kinematic regime. In particular, for forward ultraperipheral production, $pp\to p+J/\psi +p$, the LHCb experiment can reach\footnote{Note that this value corresponds to the lower limit of the $x$ interval felt by the process. In practice the main contribution to the amplitude comes from a slightly larger value of $x$,  as discussed in Section~\ref{sec:results}.}
\be
x~~\sim ~~(M_{\psi}/\sqrt{s})~e^{-Y} ~~\sim ~~3~\times ~ 10^{-6}
\ee
for $\sqrt{s}=13$ TeV and rapidity $Y=4.5$. Moreover the cross section is proportional to the square of the parton density, so the uncertainty on the PDF is reduced.

However, as mentioned in the introduction, there appear to be two disadvantages.  First, the description of the exclusive $J/\psi$ process depends on Generalized Parton Distributions (GPDs), and, second, there is a strong dependence on the choice of scale, indicating a large theoretical uncertainty.  Immediately below we note how the first disadvantage is overcome.  Then, in the next section, we discuss the removal of the sensitivity to the scale dependence.

\begin{figure} [t]
\begin{center}
\includegraphics[width=0.4\textwidth]{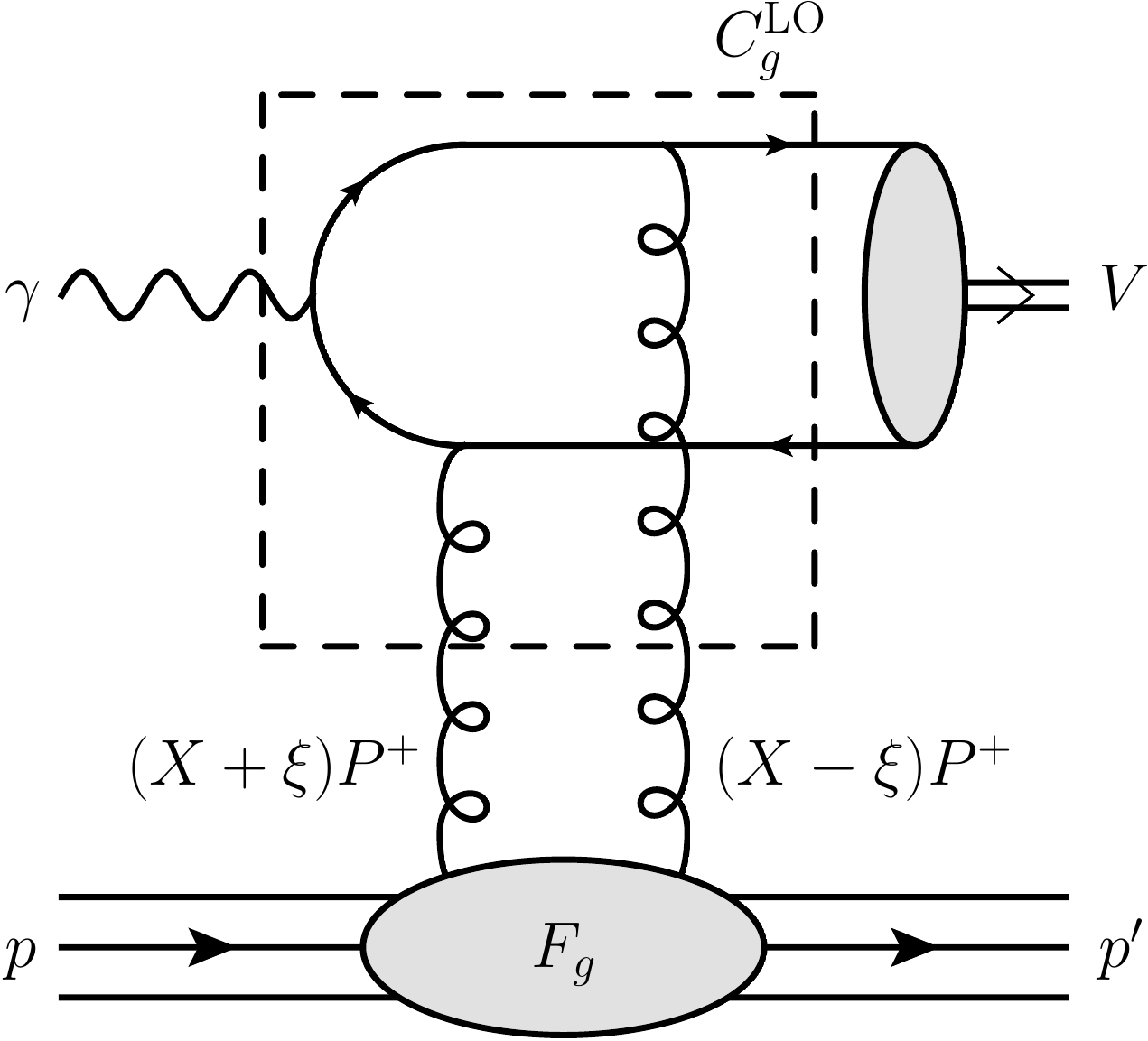}
\qquad
\includegraphics[width=0.4\textwidth]{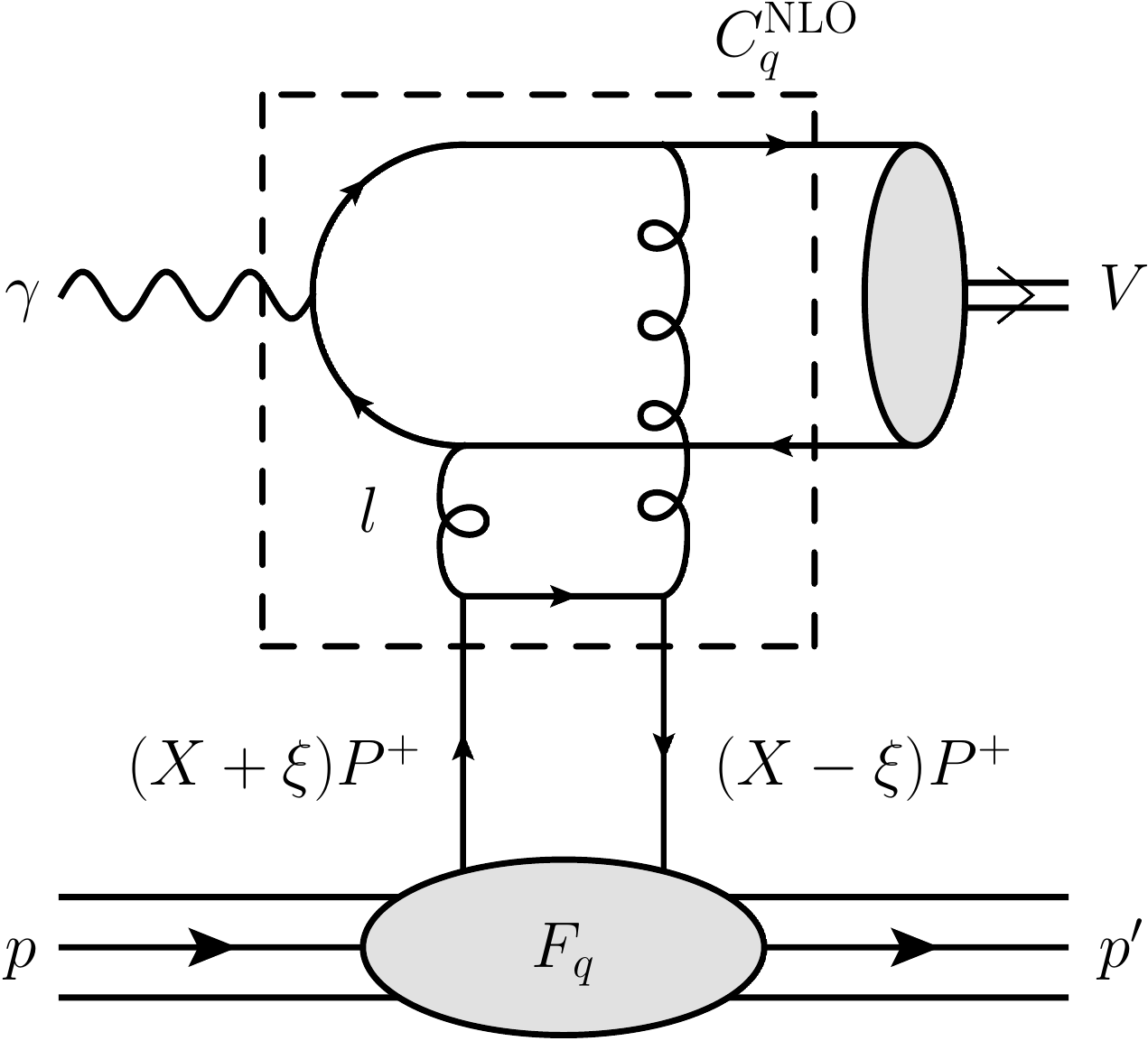}
\caption{\sf{(a) LO contribution to  $\gamma p \to V +p$. (b) NLO quark 
contribution. For  these graphs all permutations of the parton lines and couplings of
  the gluon lines to the heavy-quark pair are to be understood. Here the momentum  
  $P\equiv (p+p^\prime)/2$ and $l$ is the loop momentum. Note that the momentum fractions of the left and right partons are $x=X+\xi$ and $x'=X-\xi$ respectively; for the upper gluons we have $x' \ll x$ and so $x\simeq 2\xi$.}}
\label{fig:f2}
\end{center}
\end{figure}

Though exclusive $J/\psi$ production is described by GPDs, at very low values of $x$ and small momentum transfer $t$ the GPD can be related to the conventional integrated PDF, via the Shuvaev transform, with accuracy ${\cal O}(x)$ \cite{Shuv}. The Shuvaev transform makes use of the fact that as $\xi \to 0$ (and at $t=0$) the Gegenbauer moments\footnote{Gegenbauer moments are the analogue of Mellin moments which diagonalize the $Q^2$ evolution of PDFs. The corresponding operator diagonalizes the $Q^2$ evolution of the GPDs \cite{Or}. As $\xi \rightarrow 0$ the Gegenbauer moments become equal to the Mellin moments.} of the GPD become equal to the known Mellin moments of the PDF. Due to the polynomial condition (see e.g.~\cite{Ji}) even for $\xi \ne 0$ the Gegenbauer moments can be obtained from the Mellin moments to ${\cal O}(\xi)$ accuracy. Thus it is possible to obtain the full GPD function at small $\xi$ from its known moments. Based on this fact we can obtain an expression which transforms the low $x$ PDF to the corresponding GPD~\cite{Shuv}.

The GPD function (denoted by $F_a(X,\xi)$ with $a=g,q$ in Fig.~\ref{fig:f2}) accounts for the fact that the momenta of the `left' and `right' partons in the diagrams of Fig.~\ref{fig:f2} are different. In particular, they carry proton momentum fractions $X+\xi$ and $X-\xi$ respectively. The Shuvaev transform relates the GPD  $F_a(X,\xi)$ to the PDF$(X+\xi)$. We systematically construct GPD grids from a three-dimensional parameter space in $X, \xi/X$ and $Q^2$ with forward PDF grids taken from the LHAPDF interface \cite{LHAPDF}. It turns out that the values of $X$ that are most relevant
in the convolution of the GPD with the coefficient function are of the order of
$\xi$. Thus,
indeed, in this way we probe the gluon PDF at values of $x$ close to $2\xi$.

Strictly speaking, by using such a transform we assume that the amplitude has no additional singularities in the right half ($j>1)$  of the complex angular momentum $j$ plane.  This assumption is well motivated physically, and moreover it was shown \cite{Nockles} that the results agree with those obtained \cite{Kum} in an independent global GPD analyses of the available data.

\section{Overcoming the strong scale dependence}
\label{sec:scale}

The strong sensitivity  to the choice of scale in the  predictions for diffractive $J/\psi$ photoproduction was first observed in \cite{Ivan,JMRT3} and recently confirmed in \cite{recent}.  There are two sources for this sensitivity to the scale choice. Firstly, there is the double logarithmic contribution which contains a large ln$(1/x)$ factor.  For the region of interest, $x\sim 10^{-5}$, this means an order of magnitude enhancement. Secondly, there is double counting in the coefficient functions for $Q^2<Q^2_0$. We discuss how these problems are overcome in turn.

\subsection{Treatment of double log contributions}
It was shown in Ref.~\cite{JMRT3} that it is possible to find a scale 
(namely $\mu_F=M_{\psi}/2$) which effectively resums all the double logarithmic corrections 
enhanced by large values of ${\rm ln}(1/\xi)$ into the gluon and quark 
PDFs, where $\xi$ is the skewedness parameter of the Generalised Parton 
Distributions (GPDs).  In terms of the usual (unskewed) PDFs related to GPDs via the Shuvaev transform, $x \simeq 2\xi$.  
That is, it is possible to take the $(\alpha_S{\rm ln}(1/\xi){\rm 
ln}(\mu_F^2))$ term from the NLO gluon (and quark) coefficient functions 
and to move it to the LO GPDs.  This allows a resummation of all the 
double logarithmic, i.e  $(\alpha_S{\rm ln}(1/\xi){\rm ln}(\mu_F^2))^n$, terms in the LO 
contribution by choosing the factorization scale to be $\mu_F=M_{\psi}/2$.
The details are given in Ref.~\cite{JMRT3}, see also Ref.~\cite{DY}.

The result is that the $\gamma p\to J/\psi~p$ amplitudes, taken at factorization scale $\mu_f$,  are schematically of 
the form
\be
\label{2}
A(\mu_f)~=~C^{\rm LO} \otimes {\rm GPD}(\mu_F)~+~C^{\rm NLO}_{\rm 
rem}(\mu_F)\otimes{\rm GPD}(\mu_f).
\ee
With the choice $\mu_F=\mu_0=M_{\psi}/2$, the remaining NLO coefficient function, $C^{\rm NLO}_{\rm 
rem}(\mu_F)$, does not contain terms enhanced by ln$(1/x)\simeq\ \ln(1/\xi)$.

Thus to summarize, eq.~(\ref{2}) allows us to consider different factorization scales $\mu_f$. However the scale in the first term on the right-hand-side is fixed to be $\mu_F=m_c$ independent of the value of $\mu_f$. Since the  contribution from the second term is small we predominantly probe the gluon distribution at scale $\mu_F=\mu_0$.

\begin{figure} [t]
\begin{center}
\includegraphics[width=0.468\textwidth]{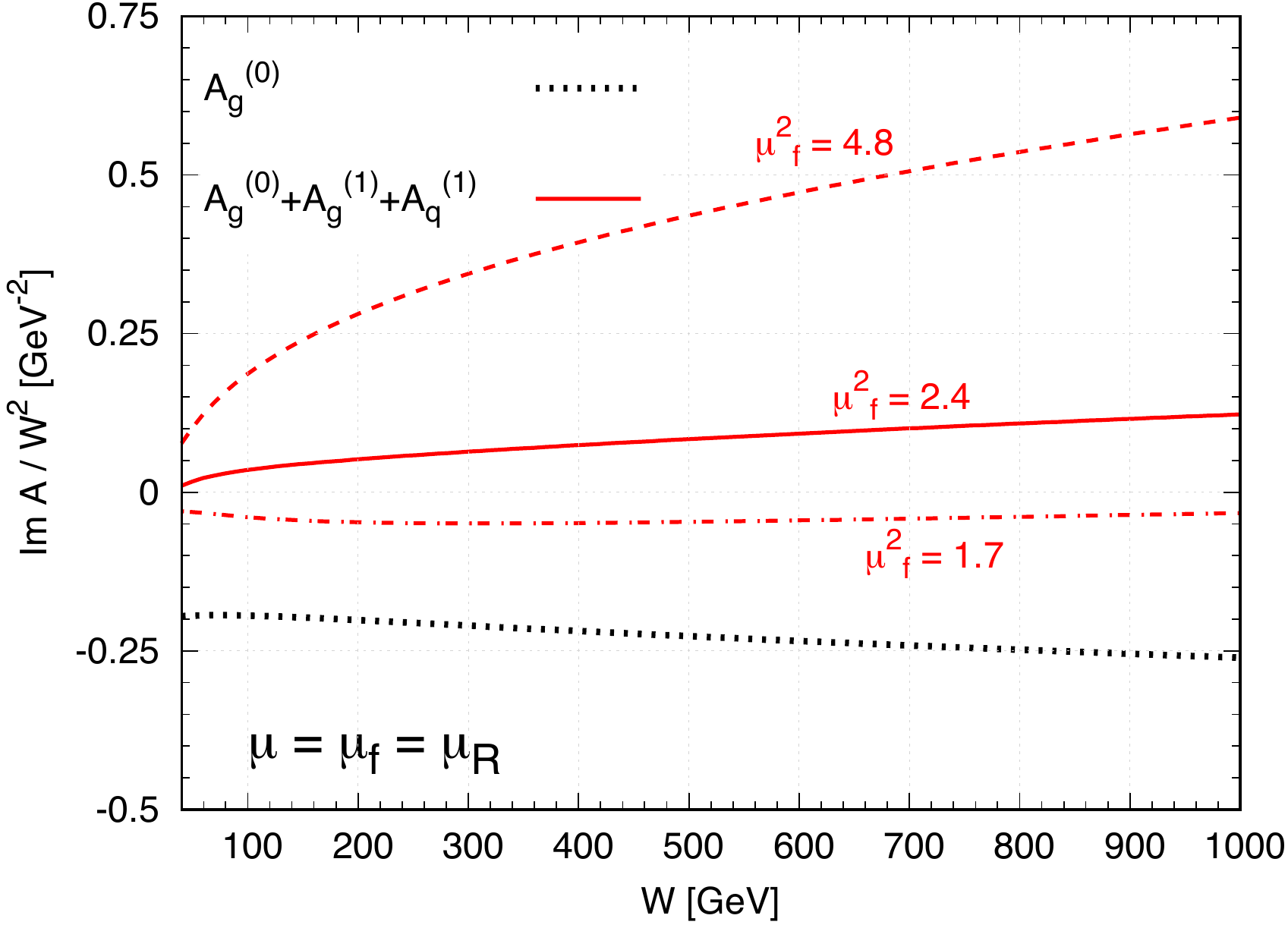}
\qquad
\includegraphics[width=0.475\textwidth]{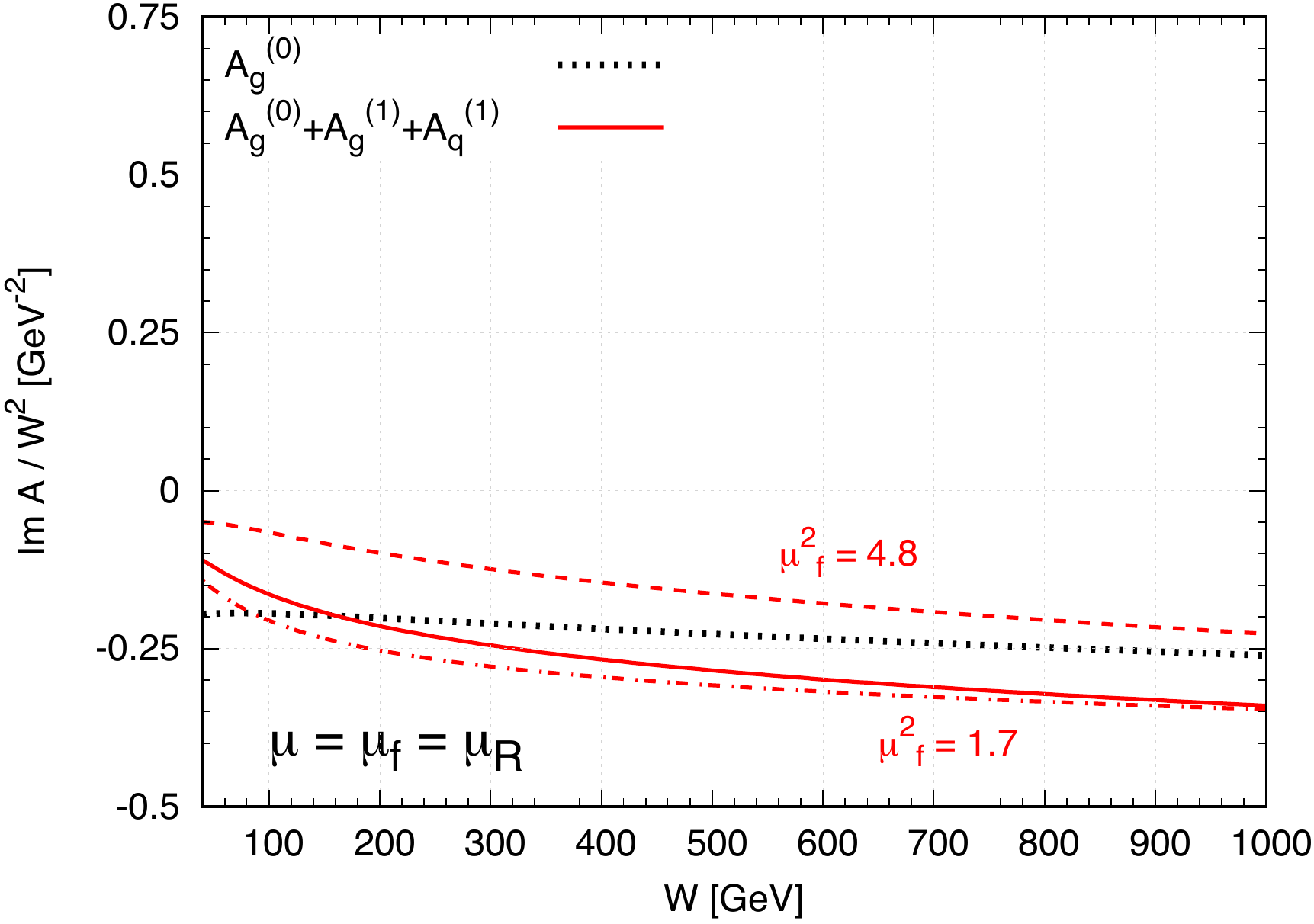}
\caption{\sf{LO and LO+NLO contributions to the imaginary part of the  $\gamma p \to V +p$ amplitude as a function of the $\gamma p$ centre-of-mass energy, $W$, with $\mu_F = m_c$ before (left panel) and after (right panel) the double counting correction has been implemented, as explained in the text. The dashed, continuous and dot-dashed (red) curves correspond to three choices of the factorization scale $\mu_f$: namely $\mu_f^2=2m_c^2,~m_c^2,~Q_0^2,$ respectively, where $m_c^2=M^2_\psi /4=2.4$ GeV$^2$.  Here $Q_0=1.3$ GeV is the starting scale of the input PDFs from CTEQ6.6 \cite{CTEQ6.6} which were used. The dotted black curve is the LO contribution.} }
\label{f3}
\end{center}
\end{figure}

Moreover, we find that after the scale $\mu_F$ in (\ref{2}) is fixed to $\mu_F=\mu_0$,  the result (shown in the left panel in Fig.~\ref{f3}) becomes more stable with respect to variations of the factorization scale $\mu_f$ in comparison to the huge variations seen in \cite{Ivan}. However note that the NLO correction is still comparable to the LO term and opposite in sign. As we discuss in  Section \ref{sec:3.2}, this is due to double counting between the NLO coefficient function and the contribution coming from DGLAP evolution. Once we
avoid this double counting, we will see that the perturbative treatment is brought under control and also that we have a further reduction of the scale sensitivity.

\subsubsection{BFKL resummation}

The possibility exists of resumming the $\alpha_s\ln(1/x)$ BFKL terms in the coefficient functions. In particular in \cite{Ivanov2}, instead of fixing $\mu_F=\mu_0$, it was proposed to resum the BFKL corrections, like $\alpha_s\ln(1/x)$, already in the coefficient function. It was stated that this would allow good scale stability to be obtained.

However, we do not resum the BFKL corrections for the following reasons.
First, we note that we cannot use the standard LO BFKL summation. We would have to account for the effects of the $Q_0$ subtraction. Also recall that LO BFKL gives the behaviour $xg\sim x^{-\omega_0}$ where
\be
\omega_0~=~(3\alpha_s/\pi)~4\ln2~\simeq~0.6
\ee
which is too large and inconsistent with the LHCb data.
Next, a detailed study \cite{Kol,KLRS} has found that at low $Q^2$ the higher-twist effects (that is, gluon reggeization and
absorptive corrections) strongly modify the low $x$ behaviour of the BFKL amplitude.
That is why the effective Pomeron intercept, measured for example, via
the vector meson diffractive photo(electro)production falls  from
$\alpha(0)_P\simeq 1+0.3$ (at large $Q^2$) down to 1+0.1 (at low $Q^2$).
  Without the BFKL resummation all these effects are absorbed
in the behaviour of the `input' phenomenological gluons.

In addition to the problems above, if the coefficient functions were to absorb the BFKL effects, then the convolution of the GPD with the coefficient function 
\be
{\rm Im} A\,(\xi)~~ \sim ~~\int_{-1}^1 \frac{\text{d}X}{X}~ C_a (X,\xi) F_a(X,\xi), \qquad (a=q,g)
\ee
  is such that the coefficient function, $C_a(X, \xi)$, occupies almost the whole available ln$(1/X)$ interval; that is the dominant contribution comes from $X\sim{\cal O}(1)$ and not $X \sim \xi$.
Thus, we would lose the main advantage of probing the unexplored very small $x$ regime.

\subsection{Treatment of double counting power corrections  \label{sec:3.2}} 
Next we consider a power correction which may further reduce the NLO 
contribution and, moreover, may reduce the sensitivity to the choice of 
scale. The correction is ${\cal O}(Q_0^2/M^2_\psi)$ where $Q_0$ denotes 
the input scale in the parton evolution which turns out to be important for 
the relatively light charm quark, $m_c \simeq M_\psi /2$. Let us explain 
the origin of this `$Q_0$ correction' following Ref.~\cite{JMRT2}.  We begin with the collinear 
factorization approach at LO. Here, we never consider parton distributions 
at low virtualities, that is for $Q^2<Q_0^2$.  We start the PDF evolution 
from some phenomenological PDF input at $Q^2=Q_0^2$. In other words, the 
contribution from  $|l^2|<Q^2_0~$  of Fig.~\ref{fig:f2}(b) (which can be 
considered as the LO diagram, Fig.~\ref{fig:f2}(a), supplemented by one 
step of the DGLAP evolution from quark to gluon, $P_{gq}$)  is already 
included in the input gluon GPD at $Q_0$.
  That is, to avoid double counting, we must exclude from the NLO diagram 
the contribution coming from virtualities less than $Q_0^2$. At large 
scales, $Q^2\gg Q^2_0$ this double-counting correction will give small 
power suppressed terms of ${\cal O}(Q_0^2/Q^2)$, since there is no 
infrared divergence in the corresponding integrals.  On the other hand, 
with $Q_0 \sim 1$ GeV and $\mu_F=m_c~ (\sim M_\psi /2$), a correction of 
${\cal O}(Q^2_0/m^2_c)$ may be crucial.

Beyond NLO single logarithmic terms, $\ln(1/x)$, may again be present 
in the amplitude. However, we anticipate that including the $Q_0$ subtraction
their impact will be much smaller.

In the present paper we use the NLO correction $C^{\rm NLO}_{\rm rem}$ for $J/\psi$  
photoproduction excluding the contribution coming from the low virtuality 
domain\footnote{Note that the value of $Q_0$ may differ from the value $q_0$ at which the initial PDFs were parametrized.  For example, in the MMHT analysis \cite{MMHT} $q_0$ is set equal to 1 GeV, but only  data with $Q^2>2$ GeV$^2$ are included in the fit.  This means that actually the input was fitted at $Q^2=2\,\text{GeV}^2$ and all the partons
below $2\, \text{GeV}^2$ are obtained by the extrapolation via the backward pure DGLAP evolution.}
$(<Q^2_0)$.  We find that for $J/\psi$ this procedure substantially 
reduces the resulting NLO contribution and, moreover, reduces the scale 
dependence of the predictions. It indicates the stability of the 
perturbative series.

Indeed, as shown in the left panel of Fig.~\ref{f3}, before the $Q_0$ subtraction the NLO corrections may exceed the value of the LO contribution and, depending on the scale, even the sign of the amplitude can change. However, after the subtraction and choosing the optimal scale $\mu_F=M_{\psi}/2$ in the leading order part of the amplitude (first term of (\ref{2})), we observe a rather good scale stability as shown in the right panel of Fig.~\ref{f3}.

\begin{figure} [t]
\begin{center}
\includegraphics[width=0.65\textwidth]{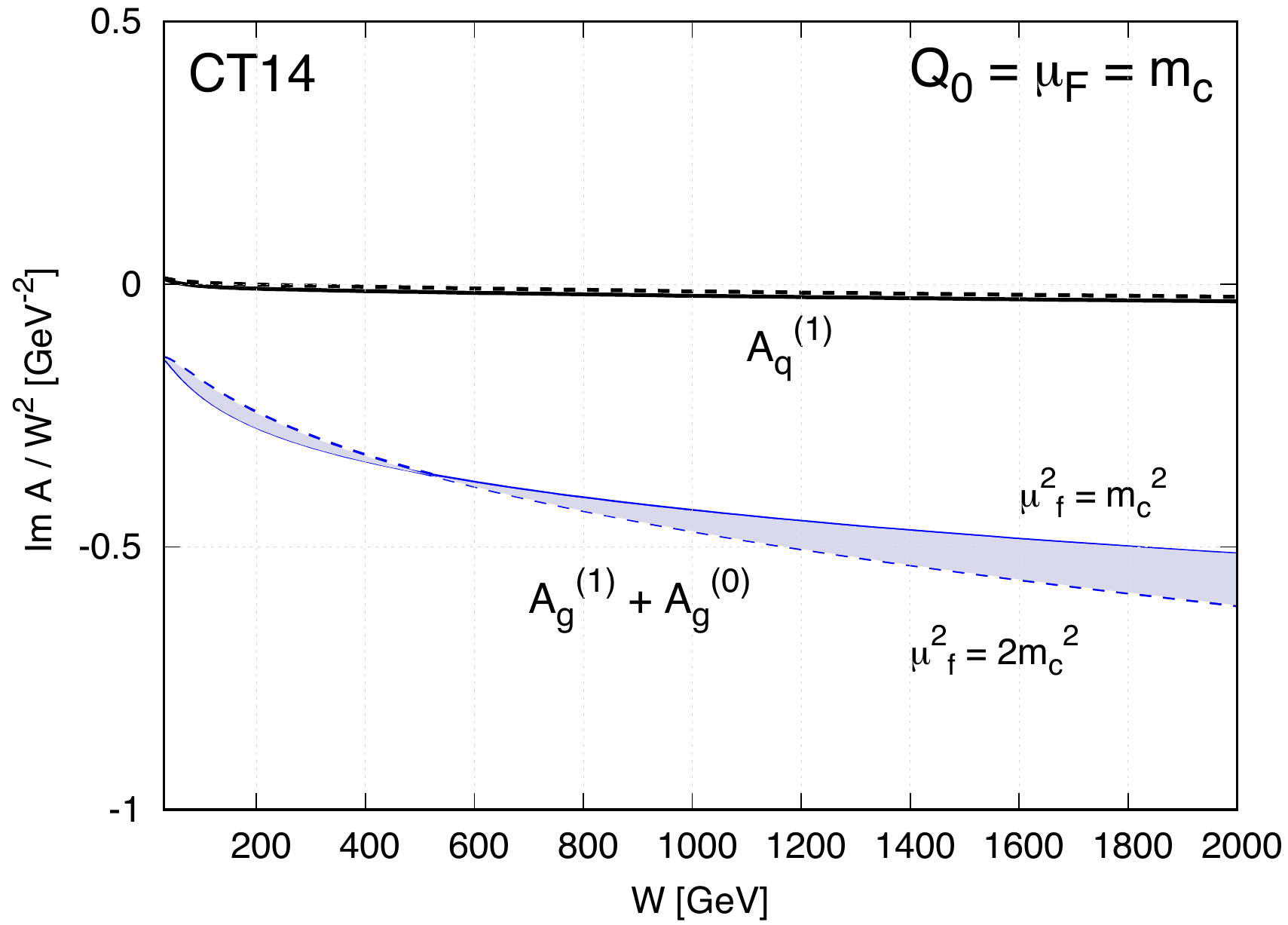}
\caption{\sf{The gluon LO+NLO and quark NLO contributions to the imaginary part of the $\gamma p \to J/\psi +p$ amplitude for two different choices of the factorization scale $\mu_f^2=\mu_R^2=m^2_c,~2m_c^2$ shown by the  continuous and dashed curves respectively. CT14 global PDFs \cite{CT14} are used and the `optimal' scale $\mu_F=m_c$ is chosen.}}
\label{f4}
\end{center}
\end{figure}

In Fig.~\ref{f4} we show the results for Im$A_a$ with $a=g,q$ for the choice $\mu_F=M_\psi/2=m_c$ for two values of the factorization scale: $\mu_f^2=m^2_c$ and $\mu_f^2=2m_c^2$. We take $\mu_R=\mu_f$.  Here $A_{a=g,q}$ are the gluon and  quark contributions to the $\gamma p\to J/\psi +p$ amplitude in the collinear factorization scheme at NLO. The plot shows the stability of the amplitude with respect to variations of $\mu_f$, and also that 
 the $Q_0$ subtraction practically fully absorbs the quark contribution. With this set-up, we can therefore say that low $x$ exclusive $J/\psi$ photoproduction probes predominantly only the gluon distribution.

\subsection{Renormalization scale}
The renormalization scale is taken to be $\mu_R=\mu_f$.
The reasons for this are as follows. Firstly, this corresponds to the BLM prescription~\cite{Brod};  such a choice eliminates the contribution proportional to $\beta_0$ (i.e. the term $\beta_0\ln(\mu^2_R/\mu^2_f)$ from the NLO terms in eq.~(3.95) of \cite{Ivan}). Secondly, following the discussion in~\cite{Luc} for the analogous QED case, we note that the new quark loop insertion into the gluon propagator appears twice in the calculation. The part with scales $\mu <\mu_f$ is generated by the virtual component ($\propto \delta(1-z)$) of the LO splitting during DGLAP evolution, while the part with scales $\mu>\mu_R$ accounts for the running $\alpha_s$ behaviour obtained after the regularization of the ultraviolet divergence. In order not to miss some contribution and/or to avoid double counting we take the renormalization scale equal to the factorization scale, $\mu_R=\mu_f$.

\section{Description of $J/\psi$ photoproduction data}
\label{sec:data}

All of the calculations presented so far are performed for the imaginary part of the production amplitude. The real part is obtained via a dispersion relation, which in the high energy limit (for the even signature amplitude) can be written in the simplified form \cite{Re}
\be
\frac{{\rm Re}A}{ {\rm Im}A}~~=~~{\rm tan}\left(\frac{\pi}{2}~\frac{\partial(\ln{\rm Im}A/W^2)}{\partial(\ln W^2)}\right).
\ee
Next we have used the nonrelativistic $J/\psi$ wave function.
   As was shown by Hoodhboy~\cite{Hood}, this provides an accuracy of a few percent, which is sufficient for our purposes.

Actually, we calculate the value of Im$A$ at $t=0$ and then restore the
total $\gamma p\to J/\psi+p$ cross section assuming an exponential $t$
behaviour with a slope
$$ B=4.9+4\alpha'_P\ln(W/W_0) ~~\mbox{GeV}^{-2}$$
with $W_0=90$ GeV and $\alpha'_P=0.06$ GeV$^{-2}$. This parametrisation grows more slowly with $W$ than the formula used by H1 \cite{H1n}, but is still compatible with the HERA data. We have chosen the slope parameter $\alpha'_P$ to be compatible with Model 4 of \cite{Diffraction} which fits a wider variety of data.

\subsection{HERA data}
\begin{figure} [t]
\begin{center}
\includegraphics[scale=0.70]{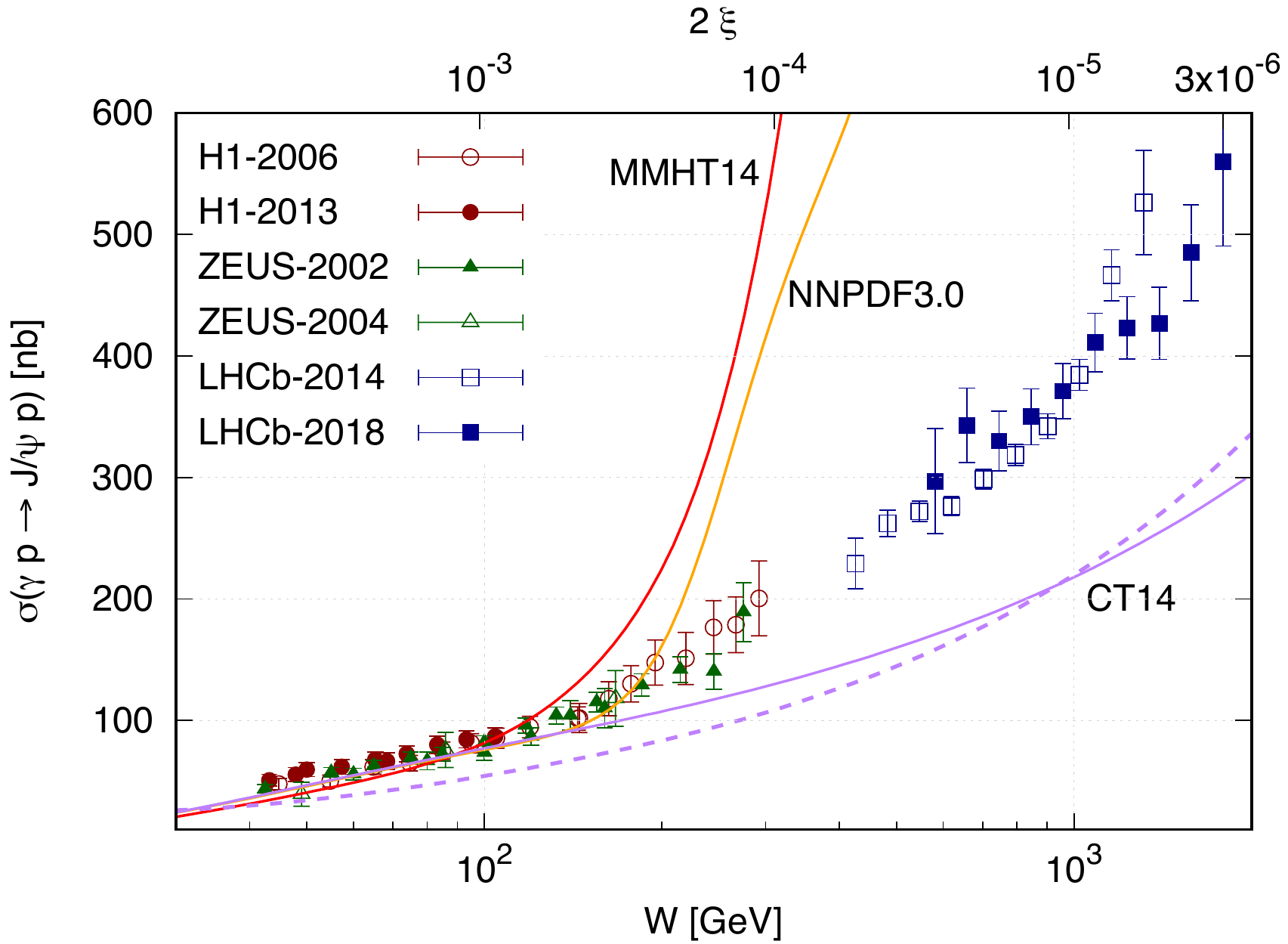}
\caption{\sf{The $\gamma p\to J/\psi+p$ data obtained at HERA \cite{HERA} and LHCb \cite{LHCb} compared with the predictions obtained using the PDFs taken from three different sets of global partons \cite{NNPDF,MMHT,CT14} with $\mu_f = m_c$ (solid lines). The dashed line for the CT14 prediction, corresponding to $\mu_f^2 = 2m_c^2$, is added to demonstrate the scale stability of our NLO predictions; but note that our optimal choice $\mu_f^2 = m_c^2$ agrees better with the HERA data.}}  
\label{f5}
\end{center}
\end{figure}

As can be seen from Fig.~\ref{f5}, the $J/\psi$ photoproduction data obtained at HERA \cite{HERA} are described reasonably well by all three sets of global partons \cite{ NNPDF, MMHT, CT14} within our collinear approach. These data sample $x$ values in the interval\footnote{We see that when $x \lapproxeq \text{few} \times
10^{-4}$ the central global partons fail to describe the HERA data.}
\be
x~~=~~M_{\psi}^2/W^2~\sim~ 10^{-3} - 10^{-4}.
\ee
In our approach we are free to choose the starting scale $Q_0$ and the $\mu_F$ scale in the NLO correction. We work at LO in NRQCD and the description used for the results shown in Fig.~\ref{f5} corresponds to the choices
\be
Q_0~=~\mu_F~=~m_c~=~M_\psi /2.
\ee
Recall that the choice $\mu_F=m_c$ provides the complete summation of the double log terms \cite{JMRT3}.
Besides giving a good description of the HERA data, the above choice of $Q_0$ and $\mu_F$ give a stable theoretical prediction also when the scales $\mu_f$ and $\mu_R$ are varied, see Figs.~\ref{f4} and 4.

\begin{figure}
\begin{center}
\includegraphics[scale=0.55]{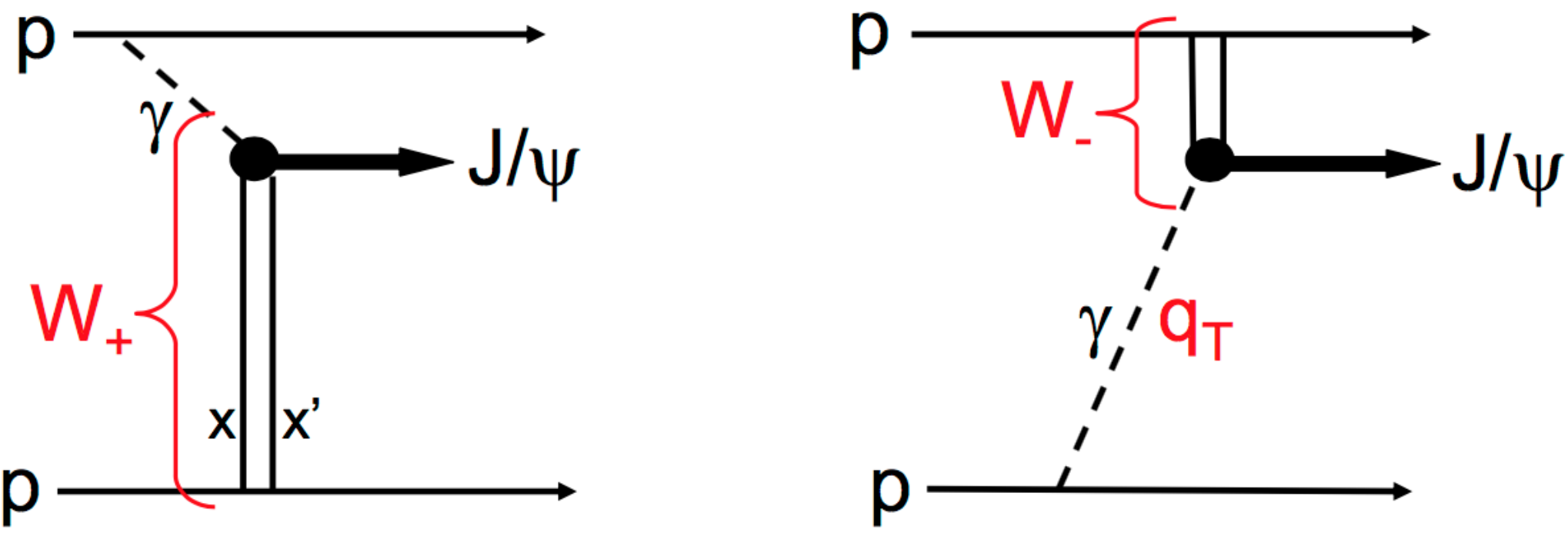}
\caption{\sf{The two diagrams describing exclusive $J/\psi$ production at the
  LHC. The vertical lines represent two-gluon exchange. Diagram (a),
  the $W_+$ component, is the major contribution to the $pp \to p+J/\psi
  +p$ cross section for a $J/\psi$ produced at large rapidity $Y$. Thus
  such data allow a probe of very low $x$ values, $x\sim M_{\psi} {\rm
    exp}(-Y)/\sqrt{s}\,$; recall that for two-gluon exchange we have
  $x\gg x'$.   The $q_T$ of the photon is very small and so the photon can be considered as a real on-mass-shell particle.}}    
\label{LHCb}
\end{center}
\end{figure}

\subsection{LHCb data}
The LHC experiments do not directly measure the cross section of {\em photoproduction}, but instead the exclusive $pp\to p~+~J/\psi~+~p$ reaction \cite{LHCb}. At small
transverse momentum of the $J/\psi$ meson this process is described by the two diagrams shown in Fig.~\ref{LHCb}. The photon can be emitted either by the upper or by the lower beam protons. Since the photon's transverse momentum, $q_T$, is much smaller than that transferred through the strong interaction amplitude (shown by the double vertical lines in Fig. \ref{LHCb}) the interference between these two diagrams is negligible. The contribution corresponding to the right graph, with a smaller 
photon-proton energy $W_-$, comes from relatively large $x$, and can be subtracted using the description of HERA
data. Thus the cross section for $J/\psi$ photoproduction at the large energy, $W_+$, may be extracted from the LHC measurements.

The last point is that in dealing with proton-proton interactions we must account for the possibility of an additional soft interaction between the two colliding protons. This interaction will generate new secondaries which  will populate the rapidity gap and  destroy the exclusivity of the event. The probability to have no such additional interaction is called the gap survival probability $S^2<1$.
The value of $S^2$ depends on the $pp$ collider energy and the partonic energy $W$. The values
of $S^2(W)$ as a function of $W$ were calculated using the eikonal model~\cite{KMR74} which well describes the data on differential $\mathrm{d}\sigma(pp)/\mathrm{d}t$ cross section and low mass diffractive dissociation. The details of the procedure to extract $\sigma(\gamma p\to J/\psi+p)$ at large $W_+$ energies is described in reference~\cite{Jones}.
Actually, in our figures we plot the low $x$ LHCb data points obtained in this way and presented 
in~\cite{LHCb}.

\section{Discussion of the results}
\label{sec:results}

The theoretical predictions, obtained by using the approach described 
above, are presented in Fig.~\ref{f5}.  There we compare our 
predictions for the cross section for $J/\psi$ photoproduction obtained using 
three different sets of global partons \cite{NNPDF,MMHT,CT14} with the 
HERA  and LHCb data.   The curves correspond to using the central values 
of the global PDFs.  At the lower energy of the 
HERA data, where the global gluon PDF uncertainty is not too large, the 
predictions agree with the experimental values reasonably well.  In the 
kinematic region covered by the LHCb experiment the present global PDF 
analyses do not sample any data, and hence they have almost no predictive 
power in this low $x$ regime.

On the other hand, as is seen from Fig. 4, by exploiting the LHCb 
data for exclusive $J/\psi$ production we have the possibility to 
greatly improve our knowledge of the gluon PDF
down to $x\sim 3\times 
10^{-6}$. The GPD$(X,\xi)$ obtained via the Shuvaev transform is driven 
dominantly by the value of $x=X+\xi \simeq 2 \xi$, while $x'=X-\xi \ll x$ is small. Recall that in the LO contribution (given by the first term 
of eq.~(\ref{2})) we sample the gluon PDF at $x=X+\xi = 2\xi$
while in the NLO contribution (the second term) the momentum fraction 
carried by the gluon may be larger.  As a check we have calculated the 
median value, $\mathrm{med}(X)$, of the corresponding $X$, defined in 
such a way that $X>\mathrm{med}(X)$ gives 0.5 of the NLO 
contribution.  In the convolution of the coefficient function with the GPD (see eq. (5)) the $X$ distribution is sharply peaked at $X\simeq \xi$ for the gluon contribution while for the quark NLO contribution the value of $ \mathrm{med}(X) \simeq 1.18\,\xi$. However, as it is seen from Fig. 3, the quark term is practically negligible.
Thus we can say that the exclusive $J/\psi$ production indeed probes the gluons at $x=X+\xi\simeq 2\xi$.


\begin{figure}[t]
\begin{center}
\includegraphics[scale=0.7]{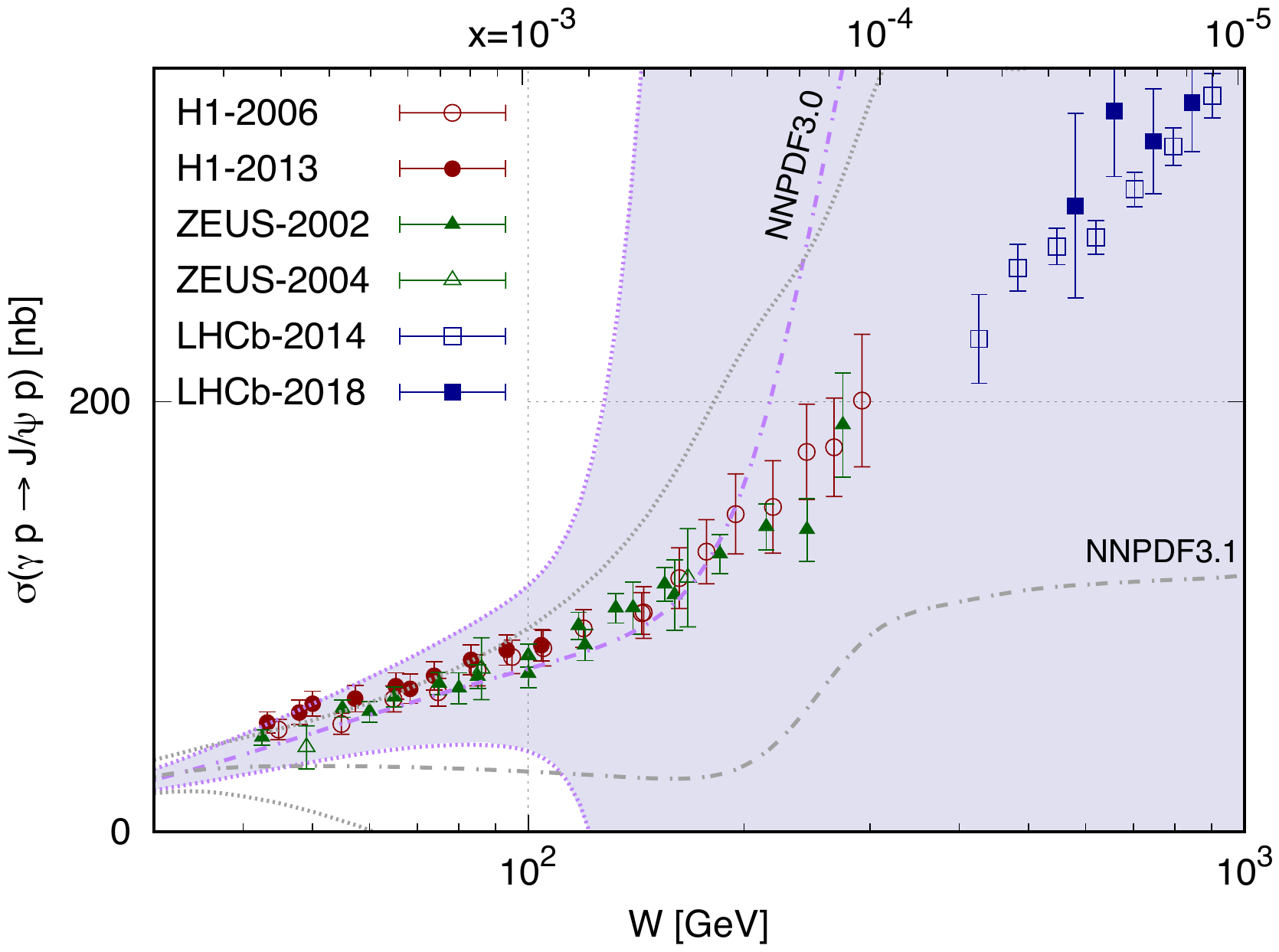}
\caption{\sf{The central scale prediction $\sigma$ for a given global input set of partons, here NNPDF3.0~\cite{NNPDF}, together with its 1$\sigma$ (shaded) error band show that the current PDF uncertainties are much greater than the experimental uncertainty and the scale variations of the theoretical result. For comparison we also show the NNPDF3.1 \cite{NNPDF31} predictions, but with the error band unshaded; in this case the $\sigma + \delta \sigma$ upper limit follows the HERA data for $x>10^{-3}$ while for smaller $x$ it widens to encompass the data. The exclusive $J/\psi$ data are therefore in a position to improve the global PDF analyses at low $x$.} }    
\label{Error}
\end{center}
\end{figure}

\section{Conclusions}
\label{sec:conclusions}

We show that the $J/\psi$ meson photoproduction process and exclusive $J/\psi$ production, $pp\to p+J/\psi +p$, at the LHC, can be consistently described in the collinear factorization framework at NLO. The choice of the optimal scale $\mu_F=\mu_0 =M_\psi /2$ effectively resums the large double logarithmic terms, i.e $(\alpha_s \ln\mu_F^2\ln(1/\xi))^n$. This, together with the $Q_0$ subtraction (needed to avoid double counting between the NLO coefficient function and the DGLAP input PDFs), leads to a largely improved scale stability of the theoretical prediction.  In other words, this framework overcomes the extremely large scale uncertainties found in the existing NLO predictions~\cite{Ivan,JMRT3,recent} of diffractive $J/\psi$ photoproduction in the collinear factorization approach.  It is not surprising that at these low scales the power correction arising from the $Q_0$ subtraction is crucial. Another power correction coming from absorptive effects should reveal itself as the saturation of the gluon density. At the moment this is not noticeable; for small $x$ the data appear to be compatible with the gluon PDF parametrization $xg\propto x^{-\lambda}$.


Huge uncertainties in the low $x$ gluon PDF found in the existing global PDF analyses reflect the fact that no corresponding low $x$ data were included in the fitting procedure. This is well illustrated in Fig. 6 which shows the prediction of, for example, the NNPDF3.0 \cite{NNPDF} parton set together with its 1$\sigma$ error band. However, using the proposed approach the good accuracy of the exclusive $J/\psi$ cross section presented by LHCb will allow the determination of the NLO gluon PDF down to $x\sim 3\times 10^{-6}$, and the HERA data will improve the determination of the gluon for $10^{-4} \lapproxeq x \lapproxeq 10^{-3}$.


\section*{Acknowledgements}

We thank Robert Thorne for a valuable discussion.  CAF thanks the CERN theory department and the IPPP at Durham University for hospitality. MGR thanks the IPPP at Durham University for hospitality. The work of TT is supported by STFC under the consolidated grants ST/P000290/1 and ST/S000879/1. 


\end{document}